# Los fact-checkers iberoamericanos frente a la COVID-19. Análisis de actividad en Facebook

The Ibero-American fact-checkers facing the COVID-19. Analysis of activity on Facebook

**Alberto Dafonte-Gómez\*, María-Isabel Míguez-González\*, Xabier Martínez-Rolán\***

*Universidade de Vigo

**RESUMEN**

**Introducción:** La pandemia de COVID-19 evidenció el papel de la infodemia como un factor agravante de la emergencia sanitaria y la relevancia del periodismo de verificación de datos en situaciones de crisis. Este artículo tiene por objetivo conocer las dinámicas de publicación e interacción en Facebook de la agrupación internacional de *fact-checking* LatamChequea Coronavirus durante los dos primeros meses (11 de marzo a 11 de mayo de 2020) tras la declaración de pandemia por COVID-19 realizada por la OMS. **Metodología:** Se realiza un análisis cuantitativo de los contenidos, metadatos y reacciones generadas en un total de 5736 *posts* publicados por 31 *fact-checkers* distintos, recopilados con Crowdtangle. **Resultados:** Entre las publicaciones más compartidas, las relacionadas con la COVID-19 generaron más reacciones y comentarios que las referidas a otras temáticas; el volumen de compartidos es elevado en relación con otras interacciones generalmente más frecuentes y las publicaciones que incluyen enlaces lograron más interacciones de todo tipo que las basadas en vídeos o imágenes. Se observan correlaciones positivas entre el número de *likes* y el número de veces que un *post* ha sido compartido, y entre la reacción de enfado y el número de comentarios. **Discusión y conclusiones:** La relevancia de los compartidos en el conjunto de interacciones indicaría un interés elevado de los usuarios por compartir los desmentidos sobre noticias falsas, algo que se distancia de investigaciones previas referenciadas. Los resultados apuntan también a la influencia de la reacción emocional expresada en la interacción con el contenido en forma de compartido o comentario.

**PALABRAS CLAVE:** COVID-19; Facebook; Infodemia; Desinformación; Verificación de datos; Periodismo digital.

## 1. Introducción: infodemia y COVID-19

El 15 de febrero de 2020 el Director General de la Organización Mundial de la Salud (OMS) afirmaba, en un discurso centrado en la COVID-19 "[…] no estamos luchando únicamente contra una epidemia; estamos luchando contra una infodemia. Las noticias falsas se propagan con más rapidez y facilidad que el propio virus, y son igual de peligrosas" (Adhanom-Ghebreyesus, 2020). Las declaraciones de Adhanom están avaladas por la evidencia científica: las noticias falsas se propagan más lejos, más rápido y a un nivel más profundo que las verdaderas, y este efecto es especialmente pronunciado cuando se abordan temáticas concretas: política, ciencia, leyendas urbanas, información financiera, terrorismo y desastres naturales (Vosoughi et al., 2018).

Si, en situaciones de normalidad, una de las principales dificultades que encuentra la ciudadanía a la hora de obtener información sobre asuntos públicos es el incremento en la circulación de noticias falsas que buscan desestabilizar a la sociedad y a las instituciones (Casero-Ripollés, 2020), en el caso de una crisis sanitaria como la de la COVID-19 la dificultad se incrementa en la misma medida en la que la sociedad demanda más información y la difusión de bulos aumenta. En relación con esta idea, Brennen et al. (2020) señalan que, en ausencia de suficiente información, la desinformación rellena los huecos en la comprensión de la audiencia, particularmente en el caso de los que desconfían de la información oficial.

Por otra parte, la velocidad de propagación de la desinformación durante una pandemia supone un riesgo adicional para la población dado que al peligro de la enfermedad se suma el peligro potencial de no saber actuar correctamente frente a ella (Zarocostas, 2020).

En Europa, según Comscore, las páginas de los medios de comunicación incrementaron notablemente su tráfico a partir de la primera semana de marzo de 2020, una subida especialmente notoria en las webs de medios locales que, en el caso español, registraron un incremento de tráfico del 158% en ese mes (Agulló, 2020). En Estados Unidos, los datos de Pew Research Center analizados por Casero-Ripollés (2020) muestran que el 92% de los ciudadanos consumieron activamente noticias sobre el COVID-19 y que el porcentaje de los que afirmaron seguir las noticias políticas "muy de cerca" se incrementó en un 32% entre la primera y la segunda quincena de marzo.

Esta demanda y consumo de información puede producir también, de forma colateral, que las posibilidades de contacto con contenidos desinformativos se incrementen, en un contexto general en el que un volumen muy importante de la población manifiesta estar preocupada por la veracidad de los contenidos al buscar información en Internet: el *Digital News Report 2020* señala que el 56% de la muestra desconfía de las noticias que encuentra en internet, una cifra que en el caso de Brasil asciende hasta el 84% y lo convierte en el "líder en desconfianza" entre los 40 países estudiados (Newman et al., 2020).

Las redes sociales son el medio de acceso a la información en el que los usuarios detectaron un mayor volumen de noticias falsas y también el entorno en el que se registró un mayor incremento durante la crisis del COVID-19 con respecto al período inmediatamente anterior (Casero-Ripollés, 2020). Con respecto a España, el estudio de Masip et al. (2020) indica que el 80,3% de los encuestados afirmó haber recibido noticias falsas entre la declaración del estado de alarma y el 10 de abril, y el *Digital News Report España 2020* (Negredo et al., 2020) señala que el 44% de usuarios de noticias online afirmó haber recibido mucha o bastante información falsa sobre la COVID-19 desde las redes sociales y específicamente de aplicaciones de mensajería como WhatsApp y Facebook Messenger.

**1.1. Desinformación y fact-checkers en tiempos de COVID-19.**

A la vista de los datos expuestos, no es de extrañar que se haya producido un enorme incremento también en las demandas de verificación de datos que los usuarios hacen llegar a los *fact-checkers*. Cristina Tardáguila, directora adjunta de la International Fact-Checking Network (IFCN), aporta datos concretos sobre este aumento a través de las declaraciones de varios profesionales: Clara Jiménez, cofundadora de *Maldita.es* señala que si en las últimas elecciones generales recibían una media de 600 solicitudes de verificación por día, durante la crisis de la COVID-19 estuvieron recibiendo entre 1500 y 2000; Joaquín Ortega, de *Newtral*, afirma haber recibido 6 veces más que en las últimas elecciones; Mehmet Atakan, de *Teyit* (Turquía) declaraba haber recibido durante el mes de marzo 7500 solicitudes de verificación, tres veces más que en el mes anterior; Matías Di Sanctis, de *Chequeado* (Argentina) recibió un 70% más de solicitudes de verificación de lo que es su media habitual, por encima de períodos de alta demanda de verificaciones, como los electorales (Tardáguila, 2020).

Según los datos recopilados por la CoronaVirusFacts Alliance auspiciada por la IFCN e integrada por más de 100 *fact-checkers* de 70 países, la media diaria de desmentidos publicados por sus miembros entre febrero y mayo fue de 138 y el período de mayor actividad se desarrolló a finales de marzo y durante el mes de abril, con un descenso paulatino pero constante durante el mes de mayo (*Fighting the Infodemic: The #CoronaVirusFacts Alliance*, 2020)

En relación con la diversidad de desinformación que circuló en lo primeros meses de pandemia, Tardáguila identifica, incluso, las temáticas predominantes con 5 oleadas de desinformación durante la crisis de la COVID-19 (Suárez, 2020), a medida que los hechos se van desencadenando: la primera se refiere a los orígenes del virus, la segunda a los vídeos de gente cayendo o directamente en el suelo en calles de China, la tercera centrada en los falsos remedios para la enfermedad o prevención del contagio, la cuarta, centrada en declaraciones de corte supremacista sobre grupos étnicos o religiosos inmunes a la enfermedad y la quinta relacionada con las pruebas, test y condiciones de cumplimiento de cuarentenas.

Según señalan Brennen et al. (2020), la mayor parte de la desinformación sobre COVID-19 se basa en la tergiversación, reformulación, descontextualización o reelaboración a partir de datos verdaderos; no abundan los contenidos completamente inventados ni hay un gran proceso de manipulación tecnológica de vídeos o fotografías, dato coincidente con los aportados por Salaverría et al. (2020). Un volumen muy relevante de estos contenidos (39%) –el más destacado de la muestra de Brennen et al. (2020)– tiene que ver con anuncios y declaraciones falsas o tergiversadas sobre políticas públicas o actuaciones gubernamentales en relación con la COVID-19 a todos los niveles, desde lo regional a la OMS o la Organización de las Naciones Unidas (ONU), algo en lo que coincide en cierta medida con la investigación de Hollowood y Mostrous (2020), que identifican 1 de cada 3 piezas desinformativas como comunicados falsos sobre acciones del gobierno con respecto a la enfermedad o sobre nuevos contagios.

**1.2. El papel de Facebook en el ecosistema de la desinformación.**

Los *Digital News Report* más recientes (Newman et al., 2018, 2019, 2020) vienen confirmando la hegemonía de Facebook como red social predominante para el acceso a las noticias y señalan el auge de WhatsApp y de los grupos privados de Facebook como fuentes de información (y desinformación) para las audiencias, una dinámica acentuada durante la crisis de la COVID-19 (Newman et al., 2020). El informe *Navigating the 'Infodemic': How People in Six Countries Access and Rate News and Information about Coronavirus* (Nielsen et al., 2020), centrado en el período de inicio de la pandemia en Europa, sitúa a Facebook como la segunda fuente de información online (sin incluir prensa) después del buscador de Google en cinco de los seis países estudiados: en Reino Unido el 33% de los encuestados afirmaba haber usado Facebook como fuente de información sobre COVID-19, en Estados Unidos el 36%, en Alemania el 25%, en España el 42% y en Argentina un 53%. La excepción es Corea del Sur, país en que la primera fuente es Naver y Facebook es la quinta, con un 20%. Es interesante hacer notar que Facebook funciona también como herramienta de mensajería privada a través de su Messenger y Grupos y que este ecosistema cerrado –encabezado ciertamente por WhatsApp– supone un entorno perfecto para la desinformación. En cuanto a la desconfianza generada por las distintas vías de acceso a la información sobre COVID-19, en España el 46% declara "no confiar" en las redes sociales y el 49% no confía en las aplicaciones de mensajería (Nielsen et al., 2020), una problemática destacada también por Magallón-Rosa (2019).

La investigación de Salaverría et al. (2020) señala que WhatsApp es la fuente más habitual de los bulos desmentidos en España por *Maldita.es*, *Newtral* y *Efe Verifica* (24,7%), mientras que Twitter (14,1%) y Facebook (4,6%) se quedan en la segunda y tercera posición, respectivamente; el 41,4% se sitúa en la categoría "sin especificar". El análisis de la base de datos publicada en portugués por Corona Verificado (2020) –que aglutina a 34 *fact-checkers* de 17 países– coincide en cierta medida con estos datos. El 10 de mayo de 2020 sumaba un total de 1523 contenidos verificados y en 1467 de ellos se indicaba el origen de dicho contenido: un 28,49% mencionan WhatsApp como fuente única o simultánea, un 20,31% mencionan Facebook de forma única o simultánea mientras que Twitter aparece citada en el 8,11%; un 20,92% se adjudican a un genérico "redes sociales", que podría incluir a alguna o todas las anteriores. Si bien WhatsApp aparece como origen principal de la desinformación en la muestra internacional –coincidiendo con lo señalado por Salaverría et al. (2020) para el caso español–, Facebook gana mucha relevancia y se alza como segunda fuente de desinformación frente a un Twitter en cifras realmente bajas.

El *Digital News Report 2020* también apunta en esta dirección: "Facebook is seen as the main channel for spreading false information almost everywhere but WhatsApp is seen as more responsible in parts of Global South such as Brazil and Malaysia" (Newman et al., 2020, p. 10). Adicionalmente, podemos señalar las investigaciones de Hollowood y Mostrous (2020), en la que los *posts* de Facebook representan el 72% de todos los casos de desinformación marcados por los fact-chechers de su estudio y Scott (2020) sobre la cantidad y el tipo de desinformación que se mueve en los grupos cerrados en Facebook relativos a la COVID-19.

**1.3. Sobre el alcance de los *fact-checkers*: la importancia de compartir.**

La variedad de la desinformación que circula en torno al COVID-19 y la diversidad de motivaciones de quienes la generan dificultan poder contraatacar con una única estrategia, pero para Brennen et al. (2020) el papel de los *fact-checkers* a este respecto es clave: "Given the importance of independent *fact-checkers*, we can only hope that more funders will be willing to support such work going forward" (p. 8). En este sentido también se pronuncian Guess et al., (2020) cuando afirman que "The most prominent journalistic response to fake news from untrustworthy websites and other forms of misleading or false information is fact-checking" (p. 476).

Las redes sociales alteran la circulación tradicional de las noticias y generan un modelo en el que la actividad de los usuarios impulsa la visibilidad de las historias (Carlson, 2016) y esto es especialmente aplicable al caso de los *fact-checkers*, que han centrado su modelo de difusión de contenidos en las redes sociales para llegar a un público más amplio y para los que la difusión orgánica de sus contenidos es crucial (Robertson et al., 2020), toda vez que, según Vargo et al. (2018) no se detecta una gran presencia del *fact-checking* en los medios de comunicación. A pesar de ello todavía no existe un gran corpus de investigación sobre los elementos y motivaciones que pueden llevar a los individuos a buscar y compartir fact-checks como práctica cotidiana (Shin & Thorson, 2017).

Para Amazeen et al. (2019) en el ecosistema actual pierden peso los medios como *gatekeepers* clásicos del periodismo y lo ganan los prescriptores: son los *gatewatchers*, usuarios con influencia sobre colectivos de dimensión variable que seleccionan, anotan y comentan noticias para sus seguidores (Bruns, 2011). La acción de compartir es valiosa para los *fact-checkers* por varios motivos: por un lado afecta no sólo al alcance, sino también al mayor interés e interacciones que generan las verificaciones que llegan a través de amigos o contactos en redes sociales (Margolin et al., 2018), lo que contribuye a lograr algunos de sus objetivos primordiales como la mejora de la educación pública, del comportamiento político y del periodismo (Amazeen, 2019); por otra parte, al igual que sucede con la prensa online (Valenzuela et al., 2017), la viralización de los contenidos de los *fact-checkers* es imprescindible desde una perspectiva empresarial para generar tráfico a los medios.

Así pues, los usuarios tienen un rol cada vez más importante para la distribución de la información en las redes a través de sus interacciones sociales como las reacciones, los compartidos y los comentarios (García-Perdomo et al., 2018). No obstante, Amazeen et al. (2019) señalan que solamente un grupo muy reducido de personas están dispuestas a compartir un *fact-check* político en redes sociales (un 11% en su estudio) y que lo hacen en mayor medida los usuarios con más edad, los más liberales (en el contexto EE.UU.) y aquellos individuos que son más activos en la búsqueda de información a través de las redes. La investigación de Amazeen et al. (2019) concluye que los individuos comparten *fact-checks*, fundamentalmente, para reforzar sus actitudes previas, un comportamiento muy relacionado con las motivaciones partidistas para compartir *fact-checks* en un contexto político descritas por Robertson et al. (2020), que puede llevar a que los seguidores de un partido compartan selectivamente mensajes de verificación de hechos que animan a su propio candidato y denigran al candidato del partido contrario, ejerciendo el papel de *gatewatchers* entre sus seguidores y los *fact-checkers* (Shin &Thorson, 2017).

Trilling et al. (2017) destacan –entre los factores que incrementan las posibilidades de que un artículo sea compartido– la referencia a asuntos locales y la proximidad geográfica y cultural del lugar en el que se desarrolla el hecho, frente a la naturaleza del conflicto o el interés humano. Encuentran también que las temáticas y enfoques positivos tienen una influencia mayor en el número de compartidos que los de temática negativa. Por su parte García-Perdomo et al. (2018) muestran que los artículos que contienen valores periodísticos como el interés humano, el conflicto y la controversia, impacto y prominencia y utilidad correlacionan positivamente con un incremento en las recomendaciones sociales en Facebook, un efecto que también se observa en temáticas como el entretenimiento, sociedad y rarezas.

El contenido es, a través de la interacción con las necesidades psicológicas de los usuarios, el único factor totalmente controlable por periodistas y medios con incidencia en las posibilidades de que un artículo sea compartido (Valenzuela et al., 2017), por lo que, además del interés y de los valores periodísticos del contenido es importante tener en cuenta que las emociones concretas que éste genera en los individuos tienen incidencia en las posibilidades de que sea compartido (Dafonte-Gómez, 2018). Si bien los contenidos que generan emociones positivas como la alegría y la sorpresa tienden a ser más compartidos, la intensidad emocional –que opera también en sentido negativo bajo la forma de rabia e ira– es también un factor destacado en las probabilidades de viralización de un contenido (Berger & Milkman, 2012; Guadagno et al., 2013; Wihbey, 2014). Aprovechando este mismo principio del comportamiento humano, redes sociales tan omnipresentes como Facebook están diseñadas para ofrecer preferentemente contenido extremo y polarizante puesto que es el que genera más engagement, más tiempo de los usuarios dentro de la plataforma y, por lo tanto, mayores beneficios (Horwitz & Seetharaman, 2020).
Investigaciones como la de dos Santos et al. (2019) señalan, además, que factores como el número de seguidores, la ratio entre número de seguidores y número de publicaciones, el uso de vídeo en los *posts* y la regularidad en la publicación afectan positivamente al acto de compartir noticias en páginas de medios en Facebook.

**1.4. *LatamChequea Coronavirus*: la alianza iberoamericana contra la infodemia.**

La CoronaVirusFacts Alliance surge en enero de 2020 en este escenario de infodemia y necesidad de información contrastada para combatirla, cuando el impacto del virus aún no había trascendido las fronteras chinas, pero la desinformación sobre él comenzaba ya a propagarse a nivel global. Esta alianza, coordinada por la IFCN, tiene como finalidad publicar, compartir y traducir el conocimiento sobre la pandemia. Es el mayor proyecto de colaboración jamás lanzado en el mundo de la verificación de datos (*Fighting the Infodemic: The #CoronaVirusFacts Alliance*, 2020).

Bajo el amparo de esta misma iniciativa surge la red LatamChequea Coronavirus en abril de 2020 integrando a 22 organizaciones de 15 países («LatamChequea Coronavirus, la plataforma para verificar información en 15 países», 2020) para divulgar en español la desinformación que circula en este ámbito geográfico (con el añadido de España); el proyecto, coordinado por *Chequeado* (Argentina) y con el apoyo de Google News Initiative va creciendo progresivamente. En el momento en el que se desarrolla la recogida de datos de esta investigación la alianza está formada por 34 organizaciones y cuenta con bases de datos en español y portugués, esta última dentro del proyecto Corona Verificado (*Aos Fatos lança base de checagens sobre Covid-19 com verificadores de Brasil e Portugal*, 2020) –apoyado también por Google News Initiative– e integrado en mayo por verificadores brasileños y portugueses que se basan en el trabajo inicial de LatamChequea Coronavirus y unen fuerzas en el proyecto.

## 2. Objetivos

A partir del marco teórico descrito, esta investigación analiza la actividad en Facebook de las iniciativas periodísticas de *fact-checking* adheridas el proyecto LatamChequea Coronavirus y plantea los siguientes objetivos:

1. Conocer el número de *posts* relacionados con la COVID-19, su volumen con respecto a otros contenidos y su frecuencia de publicación.
2. Analizar las interacciones más frecuentes con los contenidos publicados sobre COVID-19 y su relación con el tipo de publicación.
3. Determinar el nivel de interacción de los contenidos relacionados con la COVID-19 en función del número de fans de cada *fact-checker*.
4. Analizar las correlaciones entre el número de *shares* y las distintas interacciones que permite Facebook con el contenido.

## 3. Metodología

### 3.1. Materiales y métodos

El desarrollo de este trabajo emplea métodos cuantitativos y el análisis de contenido como "técnica para estudiar cualquier tipo de comunicación de una manera "objetiva" y sistemática, que cuantifica los mensajes o contenidos en categorías y subcategorías, y los somete a análisis estadístico" (Hernández-Sampieri et al., 2010, p. 260). Este método, habitual en el ámbito de las ciencias sociales, se considera idóneo para investigaciones en el área de la comunicación (Berelson, 1952), área donde más impacto ha tenido históricamente por el interés de los investigadores en averiguar las características del comunicador y su mensaje (Krippendorff, 2018).

### 3.2. Selección de la muestra.

La muestra inicial está formada por los 34 emisores, entre medios de comunicación y verificadores independientes (que no pertenecen a ningún medio), miembros de LatamChequea Coronavirus en el momento de la captura de datos (15 de mayo de 2020): *Chequeado* de Argentina; *AFP Factual* de Francia (contenido en español); *Aos Fatos*, *Estadão Verifica* y *Agência*

*Lupa* de Brasil; *Bolivia Verifica* de Bolivia; *Mala Espina Check* de Chile; *La Silla Vacía* y *ColombiaCheck* de Colombia, *#NoComaCuento* (*La Nación*) y *La Voz de Guanacaste* de Costa Rica; *Periodismo de Barrio* y *El Toque* de Cuba; *Ecuador Chequea* y *GK* de Ecuador; *Maldito Bulo* (*Maldita.es*) y *Newtral* de España; *Agencia Ocote* (*Fáctica*) de Guatemala; *Animal Político* (*El Sabueso*) y *Verificado* de México; *Despacho 505* y *La Lupa* de Nicaragua; *El Surtidor* de Paraguay; *Convoca*, *OjoPúblico*, *Verificador* (*La República*) y *Salud con Lupa* de Perú; *Observador* y *Polígrafo* de Portugal; *PoletikaRD* de República Dominicana; *UyCheck* de Uruguay; y *Cotejo.info*, *Efecto Cocuyo* y *EsPaja* de Venezuela.

Dado que la comunicación de los chequeos sobre la COVID-19 es el sujeto de estudio, la muestra se acotó temporalmente entre el 11 de marzo de 2020, día en el que la OMS declara la pandemia a nivel mundial por COVID-19, y el 11 de mayo de 2020, momento en el que se inicia la desescalada en varios países (España y Francia) o entran en nuevas fases (Alemania, Italia y Portugal) de desconfinamiento.

Del listado inicial de emisores se excluyeron *Salud con Lupa*, *GK* y *Convoca* porque durante el período analizado no presentaron ningún tipo de desmentido relacionado con la COVID-19, quedando conformada la muestra final por un total de 31 emisores.

**3.3. Extracción, selección y análisis de datos.**

Para el proceso de extracción de datos se empleó la herramienta CrowdTangle (*CrowdTangle Team*, 2020), una aplicación que rastrea la información pública en grupos públicos y páginas de Facebook, sobre todo datos de interacciones y reproducciones de vídeo. Aunque la herramienta no rastrea la totalidad de las páginas de Facebook *(CrowdTangle help,* 2020*)*, cualquier url introducida en la plataforma es añadida automáticamente a la base de datos. De los 31 emisores, solo seis (*Mala Espina Check*, *Agencia Ocote - Fáctica*, *Cotejo.info*, *Bolivia Verifica*, *Polétika RD* y *Despacho 505*) no formaban parte de la monitorización de CrowdTangle antes de realizar el estudio. Esto no afecta a la muestra, salvo por la ausencia de datos previos relativos al número de seguidores de cada página en el momento de cada publicación realizada. La integridad del volumen de interacciones está garantizada.

El proceso de extracción de datos arrojó un total de 26541 publicaciones de las que se eliminaron aquellas que no fuesen desmentidos o abordasen bulos relacionados con la COVID-19 a través de un filtrado de palabras clave utilizadas con frecuencia en las publicaciones relativas a la temática planteada (Covid, Coronavirus, SarsCov2, pandemia). Este proceso de cribado de la muestra resultó en un dataset de 5736 *posts* relacionados con la verificación en el período analizado.

Para cada publicación se consideraron las siguientes variables de análisis:
- Emisor.
- Tipo de publicación según el formato (categorización proporcionada por CrowdTangle): enlace, estatus, foto, vídeo, vídeo en vivo, vídeo nativo, YouTube.
- Número de interacciones generadas (categorización proporcionada por CrowdTangle): me gusta, me encanta, me divierte, me asombra, me enfada, me entristece, compartidos y comentarios.
- Número de fans (al inicio del período y al final del período).

Para ejecutar el análisis estadístico de los datos se empleó R. Este software se utilizó para general tablas de contingencia entre factores, distribuciones de frecuencia, medias y matrices de correlación.

A la hora de calcular la relación entre nivel de interacciones y fans se apostó por desarrollar una métrica propia: $MR_{iP}x1000fans$ = *(promedio de las ratios de interacciones por post / fans de la página en la fecha de publicación de cada post) x 1000 fans.* Esta métrica considera los datos de interacción de cada *post* en particular y los relaciona con el número de fans de la página en el momento de emisión del *post*; partiendo de esta ratio, obtenida para cada uno de los *posts* de la muestra, se halla un promedio para cada emisor que, multiplicado por 1000, aporta un dato global muy preciso de cuántas interacciones producen 1000 fans por *post*.

### 4. Resultados

#### 4.1. Emisores y *posts*.

La muestra está formada por 31 emisores, entre entidades verificadoras y medios, que generaron un total de 5736 *posts* relacionados con la verificación en el período analizado, de los cuales más de un 47% tienen que ver con el coronavirus. En términos de frecuencia, un 45,16% de los emisores publicaron menos de un *post* sobre verificación al día. *Newtral*, con 940 *posts*, y *Maldito Bulo*, con 631, son las entidades más activas en el período, con una frecuencia de publicación de más de 15 y más de 10 *posts* diarios respectivamente. Sin embargo, ha sido *Bolivia Verifica* la entidad que más ha publicado específicamente sobre COVID-19, con 358 *posts* sobre este tema, seguida de *Newtral* con 340 y *Ecuador Chequea* con 311; entre estas tres entidades suman más de un tercio de las publicaciones sobre coronavirus de la muestra. En términos porcentuales y excluyendo a *Despacho 505*, que sólo publicó dos *posts*, ambos relacionados con el coronavirus, *Ojo Público*, *Bolivia Verifica* y *Ecuador Chequea* son las entidades en las que las publicaciones sobre COVID-19 tienen un mayor peso, superior al 80% (tabla 1).

**Tabla 1.** *Nº de posts por emisor y frecuencia de publicación.*

| ENTIDAD | Total *posts* | *Posts* COVID-19 | % *posts* COVID-19 | *Posts* / día | *Posts* COVID-19 / día |
|---|---|---|---|---|---|
| **AFP Factual** | 171 | 72 | 42,11% | 2,80 | 1,18 |
| **Agência Lupa** | 211 | 138 | 65,40% | 3,46 | 2,26 |
| **Agência Pública** | 136 | 33 | 24,26% | 2,23 | 0,54 |
| **Aos Fatos** | 129 | 87 | 67,44% | 2,11 | 1,43 |
| **Bolivia Verifica** | 439 | 358 | 81,55% | 7,20 | 5,87 |
| **Chequeado** | 422 | 67 | 15,88% | 6,92 | 1,10 |
| **ColombiaCheck** | 169 | 43 | 25,44% | 2,77 | 0,70 |
| **Cotejo.info** | 26 | 14 | 53,85% | 0,43 | 0,23 |
| **Despacho 505** | 2 | 2 | 100,00% | 0,03 | 0,03 |
| **Ecuador Chequea** | 385 | 311 | 80,78% | 6,31 | 5,10 |
| **Efecto Cocuyo** | 21 | 10 | 47,62% | 0,34 | 0,16 |
| **El Sabueso** | 430 | 264 | 61,40% | 7,05 | 4,33 |
| **El Surtidor** | 20 | 10 | 50,00% | 0,33 | 0,16 |
| **El Toque** | 185 | 93 | 50,27% | 3,03 | 1,52 |
| **Es Paja** | 33 | 19 | 57,58% | 0,54 | 0,31 |
| **Estadão** | 8 | 2 | 25,00% | 0,13 | 0,03 |
| **Fáctica** | 157 | 102 | 64,97% | 2,57 | 1,67 |
| **La Lupa** | 3 | 2 | 66,67% | 0,05 | 0,03 |
| **La Silla Vacía** | 57 | 25 | 43,86% | 0,93 | 0,41 |
| **La Voz de Guanacaste** | 4 | 2 | 50,00% | 0,07 | 0,03 |
| **Mala Espina Check** | 199 | 101 | 50,75% | 3,26 | 1,66 |
| **Maldito Bulo** | 631 | 134 | 21,24% | 10,34 | 2,20 |
| **La Nación (Costa Rica)** | 47 | 7 | 14,89% | 0,77 | 0,11 |
| **Newtral** | 940 | 340 | 36,17% | 15,41 | 5,57 |
| **Observador** | 81 | 19 | 23,46% | 1,33 | 0,31 |
| **Ojo Público** | 23 | 20 | 86,96% | 0,38 | 0,33 |
| **Periodismo de Barrio** | 16 | 11 | 68,75% | 0,26 | 0,18 |
| **Polétika RD** | 21 | 8 | 38,10% | 0,34 | 0,13 |
| **Polígrafo** | 438 | 188 | 42,92% | 7,18 | 3,08 |
| **UY Check** | 3 | 1 | 33,33% | 0,05 | 0,02 |
| **Verificado** | 329 | 238 | 72,34% | 5,39 | 3,90 |
| **TOTAL** | **5736** | **2721** | **47,44%** | **94,03** | **44,61** |

**Fuente:** elaboración propia.

Los *posts* sobre COVID-19 se reparten de forma bastante homogénea en el período analizado, oscilando entre los 21 y los 79 *posts* diarios. El análisis muestra descensos suaves de actividad durante los fines de semana, donde los *fact-checkers* relajan la publicación de desmentidos, acusando un descenso total los días 31 de marzo y 18 de abril. La intensidad más alta se encuentra en los días posteriores al inicio del confinamiento en varios países europeos (entre ellos España y Portugal) y, aunque con un nivel estable, a partir de finales de abril se reduce suavemente el número de publicaciones de forma general. Los picos más significativos se producen los días 20 y 24 de marzo; sin embargo, un análisis de contenido detallado de estos *posts* no revela coincidencias de contenido destacables que pudieran haber motivado el incremento de *posts* en esas jornadas específicamente (gráfico 1).

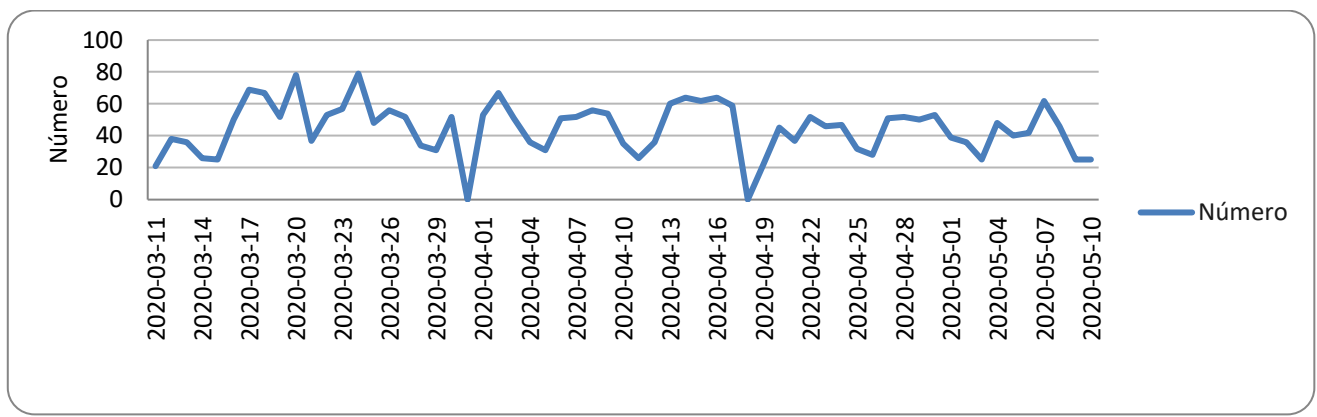

**Gráfico 1:** Número de *posts* relacionados con el coronavirus por fecha de publicación.
**Fuente:** Elaboración propia.

El análisis de fecha de publicación en función de otras variables como el emisor o el tipo de publicación según su formato no aporta datos significativos.

### 4.2. Tipo de publicación.

Los *posts* cuyo elemento principal es un enlace y los que están basados en imágenes suman en torno a un 90% de las publicaciones, tanto en el conjunto total de *posts* como en los *posts* que se relacionan con el coronavirus, si bien los *posts* basados en imagen ganan una presencia un poco mayor en este último caso. En los *posts* específicos sobre coronavirus, la publicación centrada en un enlace es la más utilizada por la mayoría de las entidades, salvo *Bolivia Verifica* y *Ecuador Chequea* que emplean fundamentalmente las imágenes y cuyo uso de los enlaces es residual (tabla 2).

**Tabla 2.** *Post por tipo de publicación.*

|  | Total *posts* |  | *Posts* COVID-19 |  |
|---|---|---|---|---|
|  | n | % | n | % |
| **Enlace** | 3436 | 59,90% | 1388 | 51,01% |
| **Estatus** | 49 | 0,85% | 27 | 0,99% |
| **Foto** | 1873 | 32,65% | 1083 | 39,80% |
| **Vídeo** | 3 | 0,05% | 1 | 0,04% |
| **Vídeo en vivo** | 7 | 0,12% | 1 | 0,04% |
| **Vídeo nativo** | 362 | 6,31% | 218 | 8,01% |
| **YouTube** | 6 | 0,10% | 3 | 0,11% |
| **TOTAL** | 5736 | 100% | 2721 | 100% |

**Fuente:** elaboración propia.

**4.3. Resultados de interacción y reacciones de los *posts*.**

En las publicaciones sobre coronavirus, los *likes*, presentes en un 97% de los *posts* y con un promedio de 70 *likes* por *post*, son la reacción más frecuente. En torno a una cuarta parte de los *posts* cuentan con reacciones negativas como la tristeza y el enfado. Casi un 89% de los *posts* se compartieron alguna vez, con un promedio de 69 compartidos por *post*, y un 58% de las publicaciones recibieron algún comentario. Sólo un 2% de los *posts* no cuentan con ninguna interacción (tabla 3).

Estos datos no difieren en gran medida de los obtenidos para el conjunto total de *posts*, aunque, en este caso, los compartidos, con un promedio de casi 81 por *post*, se posicionan como la interacción más frecuente, por encima de los *likes* (tabla 3).

**Tabla 3.** *Reacciones e interacciones por post.*

|  | **N *posts* con reacción** | | **% *posts* con reacción** | | **Promedio reacciones por *post*** | |
|---|---|---|---|---|---|---|
|  | Total | COVID-19 | Total | COVID-19 | Total | COVID-19 |
| **Me gusta** | 5588 | 2641 | 97,42% | 97,06% | 66,52 | 70,04 |
| **Me encanta** | 1787 | 811 | 31,15% | 29,81% | 2,44 | 1,47 |
| **Me divierte** | 2366 | 1052 | 41,25% | 38,66% | 7,87 | 7,50 |
| **Me asombra** | 2001 | 914 | 34,88% | 33,59% | 1,55 | 1,52 |
| **Me enfada** | 1871 | 801 | 32,62% | 29,44% | 6,20 | 5,32 |
| **Me entristece** | 1540 | 709 | 26,85% | 26,06% | 3,22 | 3,35 |
| **Compartidos** | 5151 | 2415 | 89,80% | 88,75% | 80,86 | 69,09 |
| **Comentarios** | 3602 | 1591 | 62,80% | 58,47% | 17,81 | 15,93 |
| **Interacciones totales** | 5634 | 2664 | 98,22% | 97,91% | 176,89 | 174,20 |

**Fuente:** elaboración propia.

Si analizamos las interacciones en función del tipo de publicación exclusivamente en los *posts* relacionados con el coronavirus, encontramos que los enlaces son el tipo de publicación que acumula un promedio mayor de *likes*, compartidos y comentarios, alcanzando, por tanto, un mayor promedio de interacción. Tras los enlaces, los vídeos son los que más destacan en cuanto a *likes*; las fotografías son las que acumulan un mayor promedio de comentarios e igualan a los *posts* de estatus por lo que respecta a los compartidos. No se ha tenido en cuenta la columna de vídeos en vivo, dado que sólo hay uno en toda la muestra (tabla 4).

**Tabla 4.** *Media de reacciones e interacciones en los posts COVID-19 por tipo de publicación.*

|  | **Enlaces** | **Estatus** | **Foto** | **Vídeo** | **Vídeo.en vivo** | **Vídeo nativo** | **YouTube** |
|---|---|---|---|---|---|---|---|
| **Me gusta** | 96,26 | 54,07 | 46,25 | 60,00 | 88,00 | 23,91 | 15,00 |
| **Me encanta** | 1,17 | 1,70 | 1,76 | 0,00 | 45,00 | 1,66 | 3,00 |
| **Me divierte** | 6,52 | 0,33 | 10,17 | 0,00 | 2,00 | 1,49 | 0,00 |
| **Me asombra** | 1,82 | 2,15 | 1,28 | 6,00 | 2,00 | 0,70 | 0,00 |
| **Me enfada** | 8,19 | 0,22 | 2,81 | 0,00 | 1,00 | 0,32 | 0,00 |
| **Me entristece** | 5,50 | 1,07 | 1,24 | 0,00 | 0,00 | 0,57 | 0,00 |
| **Compartidos** | 98,67 | 43,70 | 43,10 | 26,00 | 200,00 | 13,41 | 7,00 |
| **Comentarios** | 20,23 | 5,78 | 12,91 | 7,00 | 245,00 | 3,99 | 1,33 |
| **Media interacciones totales** | 238,35 | 109,04 | 119,52 | 99,00 | 583,00 | 46,05 | 26,33 |

**Fuente:** elaboración propia.

### 4.4. Repercusión general de las páginas e interacciones por emisor.

A la hora de analizar la repercusión general de las páginas de Facebook, hay que tener en cuenta que sólo se dispone de datos de fans de 25 de los 31 emisores. Las diferencias en el número de fans de los diferentes emisores son muy notorias; oscilan entre los más de tres millones y medio de *Estadão* y los poco más de tres mil trescientos de *UY Check* y dan muestra del diferente peso que estas entidades tienen en la red. Un tercio de estos emisores han incrementado sus fans en el período menos de un 1% o incluso los han perdido; por el contrario, destacan los casos de entidades como *Newtral*, *Polígrafo* o *AFP Factual*, que han incrementado sus fans en un 72%, 63% y 39% respectivamente. En promedio, los fans se han incrementado cerca de un 11% en el período analizado (tabla 5).

El cálculo de la ratio entre las interacciones por *post* por cada 1000 fans en los *posts* relacionados con el coronavirus, atendiendo al momento de publicación de cada *post* (MR$_{iP}$ por 1000 fans), permite observar que se produce una media de poco más de 3 interacciones por *post* por cada 1000 fans; sólo *AFP Factual* y *Aos Fatos* superan las 10 interacciones por *post* por cada 1000 fans (tabla 5). Considerando sólo los emisores que han publicado al menos 10 *posts* relacionados con el coronavirus en el período analizado, se observa que *Agência Lupa* obtiene los mejores resultados, con un promedio de 1160 interacciones por *post* y 6,81 interacciones por *post* por cada 1000 fans, seguida de *Aos Fatos*, con 815 interacciones de promedio y más de 13 interacciones por *post* por cada 1000 fans. *Agência Pública* supera las 500 interacciones por *post*, aunque no alcanza las tres interacciones por *post* por cada 1000 fans (tabla 5).

Cuando se comparan los datos de los *posts* relacionados con el coronavirus con los obtenidos para el conjunto total de *posts* no se obtienen diferencias significativas.

**Tabla 5.** *Relación entre interacciones por post y fans por emisor.*

| | Fans de la página | | | Total interacciones | | Promedio interacciones por *post* | | MR$_{iP}$ por 1000 fans* | |
|---|---|---|---|---|---|---|---|---|---|
| | Inicio período | Final período | Incremento (%) | Total *posts* | *Posts* COVID-19 | Total *posts* | *Posts* COVID-19 | Total *posts* | *Posts* COVID-19 |
| **AFP Factual** | 2545 | 3537 | 38,98 | 5927 | 4534 | 34,66 | 62,97 | 11,37 | 19,89 |
| **Agência Lupa** | 166639 | 173797 | 4,3 | 214215 | 160116 | 1015,24 | 1160,26 | 5,96 | 6,81 |
| **Agência Pública** | 206129 | 206897 | 0,37 | 71460 | 16785 | 525,44 | 508,64 | 2,54 | 2,46 |
| **Aos Fatos** | 59877 | 67734 | 13,12 | 79025 | 70876 | 612,60 | 814,67 | 9,83 | 13,07 |
| **Bolivia Verifica** | N/A | N/A | N/A | 58422 | 48146 | 133,08 | 134,49 | N/A | N/A |
| **Chequeado** | 81345 | 86684 | 6,56 | 43371 | 5810 | 102,77 | 86,72 | 1,23 | 1,05 |
| **Colombia Check** | 17568 | 18536 | 5,51 | 10297 | 2040 | 60,93 | 47,44 | 3,37 | 2,63 |
| **Cotejo.info** | N/A | N/A | N/A | 31 | 14 | 1,19 | 1,00 | N/A | N/A |
| **Despacho 505*** | N/A | N/A | N/A | 36 | 36 | 18,00 | 18,00 | N/A | N/A |
| **Ecuador Chequea** | 11485 | 11972 | 4,24 | 4891 | 3920 | 12,70 | 12,60 | 1,08 | 1,08 |
| **Efecto Cocuyo** | 41531 | 41875 | 0,83 | 335 | 262 | 15,95 | 26,20 | 0,38 | 0,63 |
| **El Sabueso** | 17991 | 20931 | 16,34 | 55964 | 17384 | 130,15 | 65,85 | 6,77 | 3,28 |
| **ElSurtidor** | 80403 | 81257 | 1,06 | 3669 | 1429 | 183,45 | 142,90 | 2,27 | 1,76 |
| **El Toque** | 158432 | 158350 | -0,05 | 7748 | 4543 | 41,88 | 48,85 | 0,26 | 0,31 |
| **Es Paja** | 3539 | 3620 | 2,29 | 197 | 122 | 5,97 | 6,42 | 1,67 | 1,8 |
| **Estadão*** | 3708818 | 3713404 | 0,12 | 16943 | 5214 | 2117,88 | 2.607 | 0,57 | 0,7 |
| **Fáctica** | N/A | N/A | N/A | 5677 | 4783 | 36,16 | 46,89 | N/A | N/A |
| **La Lupa*** | 14523 | 14544 | 0,14 | 145 | 58 | 48,33 | 29,00 | 3,33 | 2 |
| **La Silla Vacía** | 250622 | 252712 | 0,83 | 8123 | 3344 | 142,51 | 133,76 | 0,57 | 0,53 |
| **La Voz de Guanacaste*** | 54046 | 54617 | 1,06 | 217 | 50 | 54,25 | 25,00 | 1,00 | 0,46 |
| **Mala Espina Check** | N/A | N/A | N/A | 3478 | 1397 | 17,48 | 13,83 | N/A | N/A |
| **Maldito Bulo** | 97546 | 111519 | 14,32 | 231182 | 29349 | 366,37 | 219,02 | 3,52 | 2,13 |
| **La Nación (Costa Rica)*** | 804874 | 812057 | 0,89 | 5558 | 123 | 118,26 | 17,57 | 0,15 | 0,02 |
| **Newtral** | 10297 | 17659 | 71,5 | 33549 | 11278 | 35,69 | 33,17 | 2,53 | 2,38 |
| **Observador** | 782025 | 797817 | 2,02 | 5479 | 682 | 67,64 | 35,89 | 0,09 | 0,05 |

| **Ojo Público** | 82608 | 87843 | 6,34 | 6742 | 6679 | 293,13 | 333,95 | 3,46 | 3,94 |
|---|---|---|---|---|---|---|---|---|---|
| **Periodismo de Barrio** | 9024 | 9261 | 2,63 | 353 | 227 | 22,06 | 20,64 | 2,43 | 2,27 |
| **Polétika RD\*** | N/A | N/A | N/A | 2049 | 82 | 97,57 | 10,25 | N/A | N/A |
| **Polígrafo** | 64650 | 105517 | 63,21 | 90987 | 38977 | 207,73 | 207,32 | 2,12 | 2,12 |
| **UY Check\*** | 3314 | 3307 | -0,21 | 117 | 17 | 39,00 | 17,00 | 11,78 | 5,14 |
| **Verificado** | 30094 | 35562 | 18,17 | 48475 | 35731 | 147,34 | 150,13 | 4,49 | 4,62 |

*\*Entidades que han publicado menos de 10 posts sobre coronavirus en el período analizado.*

**Fuente:** elaboración propia.

Por último, cabe apuntar que la correlación entre el número de *posts* publicados por cada emisor y el total de interacciones es positiva, aunque no muy significativa (r=0,49 y r=0,34 respectivamente). Lo mismo sucede en la correlación entre *shares* y número de *posts* o frecuencia de publicación (r=0,48 para el total de *posts* y r=0,31 para los *posts* COVID-19).

**4.5. Correlación entre diferentes tipos de interacción.**

Si se analiza la correlación entre comentarios y compartidos con las diferentes reacciones en el caso de los *posts* COVID-19, se observa que esta es positiva en todos los casos, aunque no siempre significativa. Las correlaciones más elevadas se producen entre los compartidos y los *likes* (r=0,80) y entre los comentarios y la reacción de enfado (r=0,65). Podría deducirse, por tanto, una mayor tendencia a compartir los *posts* que gustan y a comentar los *posts* que enfadan. También se da una correlación relativamente elevada entre los comentarios y los compartidos (r=0,67) (tabla 6).

Si se comparan estos datos con los de todos los *posts*, se observa que la correlación entre compartidos y la mayoría de las reacciones es superior en el caso de los *posts* COVID-19 que en el conjunto de *posts*. Sucede lo mismo en el caso de las correlaciones entre comentarios y reacciones, salvo en el caso de los *likes* (tabla 6).

**Tabla 6.** *Correlación entre elementos de interacción y reacciones.*

|  | **Total posts** | | **Posts COVID-19** | |
|---|---|---|---|---|
|  | Comentarios | Compartidos | Comentarios | Compartidos |
| **Comentarios** | 1,00 | 0,41 | 1,00 | **0,67** |
| **Compartidos** | 0,41 | 1,00 | **0,67** | 1,00 |
| **Me gusta** | **0,59** | **0,50** | 0,58 | **0,80** |
| **Me encanta** | 0,29 | 0,18 | 0,31 | 0,33 |
| **Me divierte** | 0,56 | 0,26 | 0,62 | 0,22 |
| **Me asombra** | 0,43 | 0,27 | 0,45 | 0,44 |
| **Me enfada** | 0,55 | 0,39 | **0,65** | 0,63 |

| Me entristece | 0,37 | 0,28 | 0,47 | 0,60 |
|---|---|---|---|---|

**Fuente:** elaboración propia.

**4.6. Análisis de los *posts* más compartidos.**

Si se analizan sólo los *posts* que han sido compartidos en más de 1.000 ocasiones, se obtiene un total de 67 *posts*, de los que 30 (un 44,78%) tienen que ver con el coronavirus. Todos ellos contienen links o imágenes, si bien la proporción que representan estas últimas en los *posts* específicos sobre coronavirus es mayor que la que se da en el conjunto de *posts* (tabla 7).

En cuanto a las entidades emisoras, en el caso de los *posts* COVID-19 más compartidos, un 60% fueron emitidos por *Agência Lupa* y un 20% por *Aos Fatos*. El resto se reparten entre *Bolivia Verifica* (2 *posts*) y *Estadão*, *Maldito Bulo* y *Verificado*, con un *post* cada uno. *Maldito Bulo* es la entidad que introduce más *posts* entre los más compartidos (28), pero sólo uno de ellos trata sobre el coronavirus (tabla 7).

**Tabla 7.** *Posts con más de 1.000 shares.*

|  |  | Total *posts* | COVID-19 |
|---|---|---|---|
| **N *posts*** |  | 67 | 30 |
| **N *posts* por formato** | Link | 55 | 20 |
|  | Photo | 11 | 10 |
| **N *posts* por emisor** | AgênciaLupa | 21 | 18 |
|  | Agência Pública | 5 | 1 |
|  | Aos Fatos | 6 | 6 |
|  | Bolivia Verifica | 2 | 2 |
|  | El Sabueso | 1 | 0 |
|  | Estadão | 2 | 1 |
|  | Maldito Bulo | 28 | 1 |
|  | Polígrafo | 1 | 0 |
|  | Verificado | 1 | 1 |
| **Promedio *shares*** |  | 2400 | 2222 |
| **Promedio *likes*** |  | 994,33 | 1234,57 |

| **Promedio comments** | 282,06 | 386,63 |
|---|---|---|

**Fuente:** elaboración propia.

En relación con los datos de interacción, se observa que, dentro de estos *posts* más compartidos, los *posts* específicos sobre COVID-19 alcanzan un mayor promedio de *likes* y, sobre todo, de comentarios (tabla 7), un efecto que no se apreciaba en el análisis global de reacciones desglosado en la tabla 3.

## 5. Conclusiones

La pandemia de la COVID-19 ha supuesto una reordenación de las prioridades informativas de las sociedades, que han puesto el foco en este proceso que amenaza no sólo la salud de las personas, sino también la economía de los países, y que ha relegado a un segundo plano las temáticas que habitualmente han preocupado a la ciudadanía. Este estudio ha permitido corroborar no sólo el interés informativo de todo lo relacionado con la pandemia, sino también la abundante generación de desinformación o de información dudosa que ha tenido que ser desmentida o aclarada por las diferentes entidades comprometidas con la verificación de contenidos.

Precisamente, este análisis ha constatado que en el período inmediatamente posterior a la declaración de la pandemia casi la mitad de las publicaciones de las entidades estudiadas tuvieron que ver con el coronavirus, lo que implica, no sólo una mayor actividad por parte de los *fact-checkers*, sino también una disminución de la atención hacia las noticias falsas y los bulos que se hayan producido en ese período en otros frentes (objetivo 1). Aunque las diferencias entre los emisores son notorias, las organizaciones que han mostrado un mayor interés en la temática del coronavirus son entidades verificadoras, mientras que los medios de comunicación, aunque la han incluido entre sus desmentidos, no han puesto su foco en ella, quizá porque, por su tradición informativa, han continuado atentos a las cuestiones relacionadas con la realidad política y social, más allá de la pandemia.

Con respecto al nivel de interacción generado en torno a los contenidos vinculados con la pandemia (objetivo 2), en términos generales, la distribución de las interacciones en los *posts* sobre coronavirus es similar a la habitual en Facebook, con un elevado número de *likes*, menos compartidos, muchos menos comentarios y un número residual de otras reacciones. Aunque cabría esperar que el promedio total de interacciones fuese más elevado en los *posts* relacionados con la pandemia, dado el interés informativo del tema, se observa que las publicaciones sobre coronavirus generaron en el período analizado un promedio de interacciones similar al conjunto de los *posts*. Esto significa que el comportamiento de los usuarios a la hora de interactuar en una red social como Facebook está bastante estandarizado y es poco susceptible a variaciones como consecuencia de las temáticas abordadas. Sólo en el análisis específico de los *posts* más compartidos se aprecia una mayor interacción en las publicaciones sobre COVID-19 (que gustan más y se comentan más) en comparación con el resto.

En este contexto de estandarización cabe destacar, sin embargo, la relevancia de los "compartidos" en el conjunto de publicaciones (relacionadas con COVID-19 o no). Este resultado resulta llamativo si se contrasta con otros estudios como el de Amazeen et al. (2019), que detecta que solamente un grupo muy reducido de personas están dispuestas a compartir un fact-check político en redes sociales. Este comportamiento sí se distancia de la lógica comportamental en Facebook, donde el *like* es el principal elemento de interacción, seguido a gran distancia por la compartición del contenido o el comentario sobre la propia publicación, como constatan Carah (2014) en comunicación de marcas comerciales, Kite et al.

(2016) en comunicación sobre salud, o Martínez-Rolán (2018) en comunicación política. Esto podría interpretarse como muestra de un interés elevado de los usuarios por compartir los desmentidos sobre noticias falsas y, por tanto, hacer conscientes a las personas de su entorno del fenómeno de la desinformación, lo cual resultaría un indicador muy positivo de cara a la alfabetización de los individuos en relación con este problema y un aliciente para que los *fact-checkers* continúen desarrollando su actividad en redes sociales.

Por otra parte, atendiendo a los formatos, resulta significativo apuntar que los *posts* más abundantes son aquellos en los que el enlace es el elemento principal. Aunque autores como dos Santos et al. (2019, p.7) señalan la utilización del vídeo como uno de los elementos que contribuye a que se compartan noticias en Facebook, los formatos que priorizan el enlace son, en este caso, los que generan más interacciones de todo tipo, muy por encima de los *posts* basados en vídeos y también en mayor medida que los *posts* basados en imágenes, formatos que habitualmente destacan por su capacidad para generar reacciones. En este sentido, podemos aventurar que, cuando se trata de comprobar la veracidad de las informaciones, la referencia al contenido original a través del enlace es prioritaria para motivar el interés del usuario.

La frecuencia de publicación del emisor parece afectar positivamente, en la línea de lo apuntado por dos Santos et al. (2019, p.7), al hecho de compartir contenido, aunque los datos en este sentido no son concluyentes, por lo que no se puede señalar que tenga un efecto determinante.

Por lo que respecta al número de fans, cabe diferenciar entre los medios de comunicación, cuyas páginas cuentan con un conjunto de fans amplio y consolidado, y las entidades verificadoras, que, parten, por lo general, de números más modestos, pero están en pleno proceso de expansión y crecimiento. En este sentido, el fenómeno del coronavirus sí parece haber supuesto un espaldarazo importante para algunas de estas entidades, que han visto incrementado su número de fans de forma muy importante. Este crecimiento en número de fans es coherente con lo apuntado en el marco teórico sobre el incremento del consumo y demanda de información y verificación. De todos modos, no es posible asumir que el incremento de fans se deba sólo a la situación de incertidumbre y de necesidad informativa derivada de la pandemia, por lo que cabría corroborar esta hipótesis en futuras investigaciones, con un estudio longitudinal de la evolución de fans de este tipo de páginas.

En el análisis del nivel de interacción en relación con los fans (objetivo 3), no se detectan variaciones significativas si se consideran sólo los *posts* relacionados con el coronavirus, pero sí se aprecia que esta métrica, por lo general, es más baja en los medios de comunicación que en las entidades verificadoras, probablemente porque su número de fans es más elevado y cuantos más fans se acumulan mayor probabilidad hay de que muchos de ellos sean poco activos.

Las ratios de interacción más elevadas de la muestra corresponden a *fact-checkers* de Brasil, lo que muestra que cuentan con comunidades más activas que el resto de medios y sugiere que los fans brasileños son más participativos en la lucha contra la desinformación, hipótesis que podría verse apoyada, adicionalmente, por el dato ya citado que sitúa a Brasil como el país con un mayor porcentaje de población (84%) preocupada por la veracidad de la información a la que acceden en internet (Newman et al., 2020).

Si bien el número de compartidos de un contenido puede ser más alto, en las páginas que presentan un elevado número de fans, no se han encontrado evidencias de que el número de fans o la ratio entre fans y publicaciones afecte positivamente, más allá de lo proporcional, al acto de compartir un contenido. Es más, el análisis de los *posts* más compartidos (más de 1000 veces) revela que las publicaciones sobre coronavirus más compartidas fueron emitidas por entidades verificadoras (sólo una corresponde a un medio de comunicación), aunque la mayoría de los medios de comunicación de la muestra cuentan con un número muy superior de fans al de la mayoría de estas entidades. En este sentido, podría interpretarse que los contenidos publicados por las entidades verificadoras gozan de mayor credibilidad entre los usuarios –que, por tanto, los comparten con mayor facilidad– que aquellos difundidos por medios de comunicación sobradamente conocidos cuya tendencia o línea editorial puede generar desconfianza o sensación de subjetividad.

De los datos del análisis de correlación entre interacciones (objetivo 4) en los *posts* sobre coronavirus podría deducirse una tendencia muy clara a compartir los *posts* que gustan, que no se advierte de forma tan evidente en todo el conjunto de *posts*, aunque las emociones negativas como el enfado o la tristeza también guardan una correlación positiva relevante con los compartidos. Estos resultados son consistentes con los estudios que asocian la intensidad emocional, ya sea positiva o negativa, con las probabilidades de viralización de un contenido (Berger & Milkman, 2012; Guadagno et al., 2013; Wihbey, 2014). Así mismo, también se aprecia en los *posts* COVID-19 una cierta tendencia a recibir más comentarios en aquellos que enfadan que en aquellos que generan otras reacciones, quizá porque los usuarios desean expresar su malestar con el contenido que les irrita y recurren para ello a los comentarios.

La investigación realizada muestra que los *fact-checkers* iberoamericanos agrupados en el proyecto Latam Chequea Coronavirus han reaccionado con fuerza a la propagación de desinformación en un contexto de pandemia global, pero también que, a pesar del gran esfuerzo dedicado por los *fact-checkers* a la lucha contra la desinformación, todavía queda mucho camino por recorrer en cuanto a la optimización de sus propios fans como difusores o *gatewatchers* de los contenidos verificados, algo que se demuestra vital para que este tipo de iniciativas periodísticas tengan éxito en sus objetivos tanto sociales como empresariales.

Una limitación de este estudio está en su perspectiva meramente cuantitativa. En este sentido, cabe mencionar como línea futura de investigación la realización de un análisis que permita identificar las temáticas específicas sobre las que se han generado más bulos en el contexto de la pandemia (síntomas, vías de contagio, vacunas…) y el modo, desde un punto de vista cualitativo, en que los verificadores les han dado respuesta.

## 6. Referencias